\documentclass[12pt]{article}
\usepackage{graphicx, epsfig}
\usepackage{amsmath}
\usepackage{latexsym}
\usepackage{amstext}
\usepackage{amssymb}
\usepackage{array}
\usepackage{multirow}
\usepackage{hyperref}
\usepackage{tikz}
\usetikzlibrary{external}
\tikzset{external/force remake}
\newcommand{\ds}{\displaystyle}
\newcommand{\beq}{\begin{eqnarray}}
\newcommand{\eeq}{\end{eqnarray}}
\newcommand{\beqq}{\begin{eqnarray*}}
\newcommand{\eeqq}{\end{eqnarray*}}
\newcommand{\p}{\partial}

\newcommand{\eps}{\varepsilon}
\newcommand{\x}{\mbox{\boldmath$x$}}

\newcommand{\y}{\mbox{\boldmath$y$}}

\newcommand{\n}{\mbox{\boldmath$n$}}

\newcommand{\w}{\mbox{\boldmath$w$}}

\newcommand{\pp}{\mbox{\boldmath$p$}}

\newcommand{\sm}{\setminus}
\font\bb=msbm10 at 12pt
\def\rR{\hbox{\bb R}}

\begin{document}

\pagestyle{plain}
\begin{center}
{\large \textbf{{Mixed analytical-stochastic simulation method for the recovery of a Brownian gradient source from probability fluxes to small windows}}}\\[5mm]
U. Dobramysl\footnote{Wellcome Trust / Cancer Research UK Gurdon Institute, University of Cambridge, Tennis Court Rd, Cambridge CB2 1QN} and D. Holcman\footnote{Ecole Normale Sup\'erieure, 46 rue d'Ulm 75005 Paris, France and Mathematical Institute, University of Oxford, Woodstock Rd, Oxford OX2 6GG.}
\end{center}
\date{}
\begin{abstract}
Is it possible to recover the position of a source from the steady-state fluxes of Brownian particles to small absorbing windows located on the boundary of a domain? To address this question, we develop a numerical procedure to avoid tracking Brownian trajectories in the entire infinite space. Instead, we generate particles near the absorbing windows, computed from the analytical expression of the exit probability. When the Brownian particles are generated by a steady-state gradient at a single point, we compute asymptotically the fluxes to small absorbing holes distributed on the boundary of half-space and on a disk in two dimensions, which agree with stochastic simulations. We also derive an expression for the splitting probability between small windows using the matched asymptotic method. Finally, when there are more than two small absorbing windows, we show how to reconstruct the position of the source from the diffusion fluxes. The present approach provides a computational first principle for the mechanism of sensing a gradient of diffusing particles, a ubiquitous problem in cell biology.
\end{abstract}

\section{Introduction}
Recovering the source location from incomplete information about the emitting signal is a generic problem in several fields of science, such as finding an emitter in signal processing, the food source by smelling a few molecules and many more. In the context of cell biology, how a cell can sample its environment and decide its final destination remains open, but it starts with the detection of an external gradient concentration that the cell  must use to transform cell positional information into its genetic specialization and differentiation \cite{Wolpert,Kastakkin2007}. \\
During axonal growth and guidance, the growth cone (which is the tip of a neuronal cell) uses external concentration gradients \cite{Goodhill1,reviewREingruber} to decide whether to continue moving or to stop, to turn right or left. Bacteria and spermatozoa can orient themselves in various chemotactical or mechanical gradients \cite{Heinrich3,Heinrich5}.  However, most models in the current literature that are concerned with addressing these questions rely on computing the flux to an absorbing or reflecting ball~\cite{BergPurcell}, an absorbing or permeable ball~\cite{Endres2008,Endres2016}, or a single receptor sphere~\cite{Kaizu2014}, all of which is insufficient to differentiate between concentrations to the left or right of the cell. To enable sensing of this difference, the detectors, modeled here as small absorbing windows, should be considered individually. \\
We compute here in the first part the steady-state fluxes of Brownian particles to small absorbing windows located on the boundary of a an infinite domain. Computing the fluxes of Brownian particles moving inside a bounded domain to small absorbing windows located on a boundary falls into  the narrow escape problems \cite{Ward1,SIAM,holcmanschuss2013,holcmanschuss2015,Ward2,Ward3}  and has also been studied numerically ~\cite{Lakhani2017}. However, the mean passage time to a small hole becomes infinite in an unbounded domain due to long excursions to infinity of Brownian trajectories. This difficulty is resolved here by computing the flux directly using two methods: first, we compute the flux of Brownian particles to small absorbers located on the half-plane, a disk in $\rR^2$ and in a narrow band. The asymptotic computations are obtained by matched asymptotics of Laplace's equation in infinite domains. \\
In the second part, we develop a mixed numerical procedure to avoid tracking Brownian trajectories in the entire infinite space. We generate particles near the absorbing windows, computed from the analytical expression of the exit probability on an artificial boundary without introducing any artifacts \cite{Nadler1,Nadler2}. This method avoids the costly computation of particle trajectories in the unbounded environment (e.g. extracellular space in the brain or cells moving in two dimensional chamber), containing large excursions away from the cell, thereby allowing direct simulations of Brownian trajectories in the region of interest close to the cell. In the absence of such procedure, these simulations would be next to impossible due to the aforementioned infinite mean passage time.\\
We show that the results of both independent methods (Asymptotic and numerical) agree. The local geometry and distribution of  windows does matter for the reconstruction of the source position: we show that it is indeed  possible to recover the source of a gradient already with three receptors. Finally, the location of the windows might also be critical for the sensitivity of detection: for example, the flux of Brownian particles to small targets depends crucially on their localization \cite{holcmanschuss2013,JPA-holes,Ward2,Ward3,holcmanschuss2015,Ward1}. In summary, the manuscript is organized as follows. First, we compute asymptotically the flux of Brownian particles to receptors. Second, we introduce the mixed simulation method. In the third part, we present several applications to various geometry: half-space, a disk and a disk in a narrow band. In the fourth and last section, we apply the methods to reconstruction the source location.

\section{Fluxes of Brownian particles to small targets in an open space}
Brownian molecules are produced by a steady-state source located at position $x_0$ in an open space such as the two-dimensional real space $\rR^2$. The steady-state distribution of particles, $P_0$, is the solution of the Green's function
\beq
-D\Delta P_0(\x) & = & Q\delta(\x-\x_0) \;\;\text{ for }\;\; \x\, \in \,\rR^2 \label{eqDP1}
\eeq
where the parameter $Q>0$ measures the injection rate of particles. We study here the flux received by an obstacle $\Omega$ containing $N$-small absorbing windows $\p\Omega_{1}\cup\hdots\cup \p\Omega_{N}$ on its boundary $\p \Omega$. The fluxes of diffusing particles on the windows can be computed from solving the mixed boundary value problem (we set now $Q=1$) \cite{holcmanschuss2015}
\beq
-D\Delta P_0(\x) & = & \delta(\x-\x_0) \;\;\text{ for }\;\; \x\, \in \,\rR^2\sm \Omega \label{eqDP2}\\
 \nonumber \ds \frac{\p P_0}{\p n}(\x) & = & 0 \;\;\text{ for }\;\; \x\, \in\, \p\Omega \sm (\p \Omega_{1}\cup\hdots \cup\p\Omega_{N})\\
 \nonumber P_0(\x) & = & 0 \;\;\text{for}\;\; \x\, \in\, \p\Omega_{1}\cup\hdots\cup \p\Omega_{N}
 \eeq
The reflecting boundary condition  accounts for the impenetrable walls and diffusing molecules are reflected on the surface $ \p\Omega_r=\p\Omega \sm (\p\Omega_{1}\cup\hdots\cup \p\Omega_{N})$. The absorbing boundary condition on each window $\p \Omega_{1}\cup\hdots \cup\p\Omega_{N}$ represents the extreme case where the binding time of particles is fast and the particle trajectories are terminated.

Although the probability density $P_0(\x)$ diverges when $|\x|\rightarrow\infty$, we are interested in the splitting probability between windows, which is the ratio of the steady-state flux at each hole divided by the total flux through all windows:
\beq
J_k= \ds \frac{\ds \int_{\p\Omega_{k}} \ds \frac{\p P_0(\x)}{\p \n} dS_{\x}}{\ds \sum_{q} \int_{\p\Omega_{q}} \ds \frac{\p P_0(\x)}{\p \n} dS_{\x}}.
\eeq
In two-dimensions, due to the recurrent property of the Brownian motion, the probability to hit a window before going to infinity is one, thus the total flux is one:
\beq
\sum_{q} \int_{\p\Omega_{q}} \ds \frac{\p P_0(\x)}{\p \n} dS_{\x}=1.
\eeq
We shall now compute the fluxes asymptotically for three different configurations: 1 - when the windows are distributed on a line in half-plane, 2 - when there are located on a disk in the entire space, and 3 - when the disk is located in a narrow band. We use the Green-Neumann's function and the method of matched asymptotics \cite{Ward1,Ward3}.
\subsection{Fluxes to small absorbers on a half-plane} \label{sec:half-plane}
\begin{figure}[http!]
\centering
\includegraphics[scale=1]{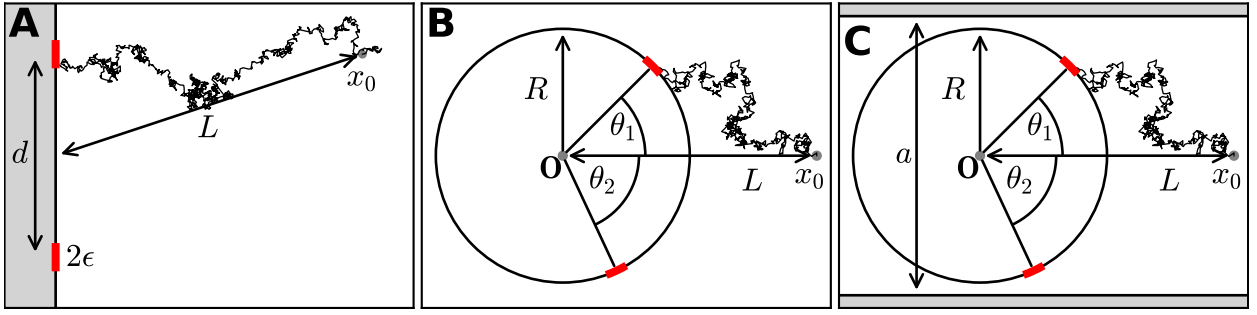}
\caption{\textbf{Brownian fluxes to small windows in different geometries}. (A) Two windows of size $2\epsilon$ are placed on the boundary of half-space a distance $d$ apart. Diffusing particles are released from a source at $\x_0$ at a distance $L=|\x_0|$ and are absorbed by one of the windows. (B) Two absorbing windows of size $2\epsilon$ are placed on the circumference of a disk with radius $R$ at angles $\theta_1$ and $\theta_2$ with the $x$-axes. As before, particles are released at the source position $\x_0$. (C) Two windows are placed on a disk as in (B), inside an infinitely long strip with reflecting walls at $y=\pm a$.}
\label{Fig1}
\end{figure}

We now estimate the fluxes of Brownian particles to two absorbing small holes, $\p \Omega_1 = \left\lbrace x=0,\;z=z_1+s\left| s \in [-\eps_1/2, \eps_1/2]\right.\right\rbrace$ and $\p \Omega_2= \left\lbrace x=0,\;z=z_2+s\left| s \in [-\eps_2/2, \eps_2/2]\right.\right\rbrace$ when the source is located at $\x_0\in\Omega$, which is the two-dimensional half-plane $\Omega = \left\lbrace  (x,z) \in \mathbb{R}^2, x>0\right\rbrace$ (Fig.~\ref{Fig1}A). Diffusing particles are reflected everywhere on the boundary of half-space, except at the two small targets.
The boundary value problem in equation \ref{eqDP2} for two windows reduces to
\beq
-D\Delta P_0(\x) & = & \delta(\x-\x_0) \;\;\text{ for }\;\; \x\, \in \,\rR_{+}^2 \label{eqDP3}\\
 \nonumber \ds \frac{\p P_0}{\p n}(\x) & = & 0 \;\;\text{for}\;\; \x\, \in\, \p\rR_{+}^2 \sm (\p \Omega_{1}\cup\p\Omega_{2})\\
 \nonumber P_0(\x) & = & 0 \;\;\text{for}\;\; \x\, \in\, \p\Omega_{1}\cup \Omega_{2}.
\eeq
We set $D=1$ and derive a solution of equation \ref{eqDP3} in the small window limit. We construct an inner and outer solution. The inner solution is constructed near each small window \cite{Ward} by scaling the arclength $s$ and the distance to the boundary $\eta$ by $\bar \eta=\frac{\eta}{\eps}$ and $\bar s=\frac{s}{\eps}$ (we use here the same size $\eps_1=\eps_2=\eps$), so that the inner problem reduces to the classical two-dimensional Laplace equation
\beq
\Delta w=0 \hbox{ in } \rR_+^{2}\\
\frac{\p w}{\p n}=0 \hbox{ for } |\bar s|>\frac{1}{2},  \bar \eta=0\\
w(\bar s, \bar \eta)=0 \hbox{ for } |\bar s|<\frac{1}{2},  \bar \eta=0.
\eeq
The far field behavior for $|\x|\rightarrow \infty$  and for each hole $i=1, 2$ is
\beq
w_i(\x) \approx  A_i \{\log|\x-\x_i|-\log \eps +o(1)\},
\eeq
where $A_i$ is the flux
\beq\label{flux}
A_i=\frac{2}{\pi} \int_0^{1/2} \frac{\p w(0,\bar s)}{\p \bar \eta}d\bar s.
\eeq
The general solution of equation \ref{eqDP2} with $n=2$ is obtained from the outer solution of the external Neumann-Green's function
\beq
-\Delta_{\x} G(\x,\y) & = & \delta(\x-\x_0) \hbox{ for } \x\, \in \,\rR_{+}^2, \label{eqDP2b}\\
\frac{\p G}{\p n_{\x}}(\x,\x_0) & = & 0\hbox{ for } \x \,   \in\, {\partial\rR_{+}^2}.
\eeq
given for $\x, \x_0 \in \rR_{+}^2$ by
\beq\label{eqDP2bs}
G(\x,\x_0)=\frac{-1}{2\pi }\left(\ln |\x-\x_0| + \ln \left|\x-\bar{\x_0}\right| \right),
\eeq
where $\bar{\x_0}$ is the symmetric image of $\x_0$ through the boundary axis $0z$. The uniform solution is the sum of inner and outer solution (Neuman-Green's function)
\beq\label{gexp}
P(\x,\x_0)=G(\x,\x_0)  +A_1 \{\log|\x-\x_1|-\log \eps\}+A_2 \{\log|\x-\x_2|-\log \eps\} +C,
\eeq
where $A_1,A_2,C$ are constants to be determined. To that purpose, we study the behavior of the solution near each point $\x_i$. In the boundary layer, we get
\beq
P(\x,\y)\approx A_i \{\log|\x-\x_i|-\log \eps\}.
\eeq
Using this condition on each window, we obtain the two conditions:
\beq\label{cond1}
G(\x_1,\x_0)  +A_2 \{\log|\x_1-\x_2|-\log \eps\} +C=0\\
\nonumber
G(\x_2,\x_0)  +A_1 \{\log|\x_2-\x_1|-\log \eps\} +C=0.
\eeq
Due to the recursion property of the Brownian motion in dimension 2, there are no fluxes at infinity, thus the conservation of flux gives:
\beq \label{compatibility}
\int_{\p\Omega_{1}} \frac{\p P(\x,\y))}{\p \n}dS_{\x} + \int_{\p\Omega_{2}} \frac{\p P(\x,\y))}{\p \n}dS_{\x} =-1.
\eeq
In the limit of two well separated windows ($|\x_1-\x_2| \gg 1$), using the condition for the flux in equation \ref{flux} we get for each window $i=1,2$
\beq
\int_{\p\Omega_{i}} \frac{\p P(\x,\y))}{\p \n}dS_{\x}= -\pi A_i
\eeq
(the minus sign is due to the outer normal orientation), thus
\beq \label{compatibility2}
\pi A_1+\pi A_2=1.
\eeq
Using relation \ref{compatibility2} and \ref{cond1}, we finally obtain the  system of two equations to solve
\beq
\frac{G(\x_1,\x_0)-G(\x_2,\x_0)}{\{\log|\x_1-\x_2|-\log \eps\}} +(A_2-A_1) =0\\
A_1+ A_2=\frac{1}{\pi}.
\eeq
The absorbing probabilities are given by
\beq \label{eq:p2-halfplane}
P_2=\pi A_2&=&\frac12+ \frac{\pi}{2}\frac{G(\x_1,\x_0)-G(\x_2,\x_0)}{\{\log|\x_1-\x_2|-\log \eps\}} \\
\label{eq:p2-halfplane2}
&=&\frac12- \frac{1}{4}\frac{\ln \frac{|\x_1-\x_0|\left|\x_1-\bar{\x_0}\right|}{|\x_2-\x_0|\left|\x_2-\bar{\x_0}\right|}}{\{\log|\x_1-\x_2|-\log \eps\}}.
\eeq
and
\beq
P_1=\frac12+\frac{1}{4}\frac{\ln \frac{|\x_1-\x_0|\left|\x_1-\bar{\x_0}\right|}{|\x_2-\x_0|\left|\x_2-\bar{\x_0}\right|}}{\{\log|\x_1-\x_2|-\log \eps\}}.
\eeq
These probabilities precisely depend on the source position $x_0$ and the relative position of the two windows. When one of the splitting probabilities (either $P_1$ or $P_2$) is known and fixed in $[0,1]$, recovering the position of the source requires inverting equation \ref{eq:p2-halfplane2}. For $P_2=\alpha \in [0,1]$, the position $\x_0$ lies on the curve
\beq\label{source}
S_{source}=\{ \x_0 \hbox{ such that} \, \frac{|\x_1-\x_0|\left|\x_1-\bar{\x_0}\right|}{|\x_2-\x_0|\left|\x_2-\bar{\x_0}\right|}=\exp \left(  (4\alpha-2)\{\log|\x_1-\x_2|-\log \eps\} \right) \}.
\eeq
At this stage, we conclude that knowing the splitting probability between two windows is not enough to recover the exact position distance of the point source $\x_0$, because it leads to a one dimensional curve solution. However the direction can be obtained by simply checking which one of the two probability is the highest.
\subsection{Fluxes to small windows on a disk} \label{sec:disk-free-space}
A similar asymptotic can be derived for the splitting probability when the domain containing the windows is a disk of radius R. The boundary condition are similar: there are no particle fluxes except on $\p \Omega \sm \left(\p\Omega_1\cup \p\Omega_2\right)$ and the two windows $\p\Omega_1\cup \p\Omega_2$ remain absorbing (Fig. \ref{Fig1}B).

We recall that the external Neumann-Green's function of a disk $D(R)$ of radius $R$, solution of the boundary value problem
\beq
-\Delta_{\x} G(\x,\y) & = & \delta(\x-\y) \hbox{ for } \x\, \in \,\rR^2 -D(R), \label{eqDP2b_disk}\\
\frac{\p G}{\p n_{\y}}(\x,\y) & = & 0\hbox{ for } \x \,   \in\, \p D(R).
\eeq
is given explicitly for $\x, \y \in \rR^2 -B(R)$ by
\beq
G_B(\x,\y)=\frac{-1}{2\pi }\left(\ln |\x-\y| + \ln \left|\frac{R^2}{|\x|^2}\x-\y\right| \right),
\eeq
It is the sum of two harmonic functions with a singularity at  $\y  \in \,\rR^2 -B(R)$  and an image singularity at $\frac{R^2}{|\y|^2}\y \in B(R)$.  A direct computation shows that $\ds \frac{\p G(\x,\y)}{\p r}|_{r=R}=0$, where $\x=r e^{i\theta}$. Following the derivation given for the half-plane above, we can use the Neumann-Green function of the disk \ref{eqDP2b_disk} directly in expression \ref{eq:p2-halfplane} and obtain the probability to be absorbed on each window: for window 2
\beq \label{eq:p2-disk}
P_2=\pi A_2&=&\frac12+ \frac{\pi}{2}\frac{G_B(\x_1,\x_0)-G_B(\x_2,\x_0)}{\{\log|\x_1-\x_2|-\log \eps\}} \\
&=&\frac12- \frac{1}{4}\frac{\ln\frac{\ds |\x_0-\x_1| \left|\ds \frac{R^2}{\ds |\x_0|^2}\x_0-\x_1\right|}{\ds |\x_0-\x_2| \left|\ds \frac{R^2}{\ds |\x_0|^2} \ds \x_0-\x_2\right|}}{\{\log|\x_1-\x_2|-\log \eps\}}.
\eeq
and for window 1
\beq \label{eq:p2-disk2}
P_1=\pi A_1 =\ds \frac12+ \frac{1}{4}\frac{\ln\frac{\ds |\x_0-\x_1| \left|\frac{R^2}{|\x_0|^2}\x_0-\x_1\right|}{\ds|\x_0-\x_2| \left|\ds\frac{R^2}{\ds |\x_0|^2}\x_0-\x_2\right|}}{\{\log|\x_1-\x_2|-\log \eps\}}.
\eeq
\subsection{Splitting fluxes with many windows} \label{manywindows}
The general solution of equation \ref{eqDP2} is given by
\beq\label{gexp_many}
P(\x,\x_0)=G(\x,\x_0)  +\sum_{k}A_k \{\log|\x-\x_k|-\log \eps\} +C,
\eeq
where $A_1,..,A_N,C$ are $N+1$ constants to be determined. We derive a matrix equation using the solution behavior near the center of the windows $\x_i$,
\beq
P(\x,\y)\approx A_i \{\log|\x-\x_i|-\log \eps\},
\eeq
and obtain the ensemble of conditions for $i=1..N$
\beq\label{cond_many}
G(\x_i,\x_0)  +\sum_{k\neq i} A_k \log\frac{|\x_i-\x_k|}{\eps} +C=0.
\eeq
The final equation is given by total flux condition:
\beq
\sum_k \int_{\p\Omega_{k}} \frac{\p P(\x,\y))}{\p \n}dS_{\x} =-1.
\eeq
When the absorbing windows are well separated compared to the distance $|\x_i-\x_j| \gg 1$, a direct computation using \ref{gexp_many} gives
\beq\label{condn}
\sum_i \pi A_i =1.
\eeq
The ensemble of conditions \ref{cond_many} and \ref{condn} is equivalent to a matrix equation
\beq\label{sys1}
[a] \mathcal{A} =\mathcal{B},
\eeq
{where for $i\neq j$, $i,j\le N$, $a_{ij}=\log\frac{|\x_i-\x_j|}{\eps}$, $a_{i,N+1}=a_{N+1,i}=1$ for $i\le N$, and $a_{ii}=0$ for $i=1..N+1$,
\beq
\mathcal{A}&=& (A_1,..,A_n,C)^T.\\
\mathcal{B}&=& (-G(\x_1,\x_0) ,..,-G(\x_n,\x_0),1/\pi)^T
\eeq
The matrix $[a]$ is symmetric and invertible, but does not have a specific structure, rendering it difficult to compute an explicit solution for a large number of windows in general. However, system \ref{sys1} can be straightforwadly solved numerically to find the unique solution $A_1,..,A_n$ and the constant $C$.}

\section{Construction of a hybrid analytical-stochastic simulations of Brownian particles to small windows}
We present a numerical method to simulate efficiently in a two-dimensional infinite domain, the splitting probability of Brownian particles to small windows located on the boundary of an {obstacle $\Omega$}. The Brownian particles are generated at a single source point $x_0$.

{It always possible to run naive Brownian trajectories, starting from the source, however, the mean arrival time of a Brownian particle to a target in a infinite two-dimensional domain is infinite, which would render the computational effort prohibitive. Therefore, naive simulations are inefficient, especially when computing average fluxes due the very large excursions of Brownian trajectories before they hit their targets. However the probability for any particle to hit a window is one, hence we dedicate the present section to develop a mixed stochastic simulations for computing the splitting probability.

To resolve the difficulties associated to naive Brownian simulations, we now introduce the simulation procedure. The goal of this procedure is to efficiently produce large ensembles of trajectories for estimating the splitting probability of Brownian particles generated at position $x_0$ and absorbed at small windows located on the surface of a two-dimensional domain.  The domains are either a disk of radius $R$ or the boundary of half-space, but any shaped is possible. The procedure can be generalized to any obstacle surface in any dimensions, where random particles evolve in an unbounded space. We now describe the mixed algorithm consisting of two steps:}

\subsection{Hybrid analytical-stochastic algorithm} \label{s:algo}
\begin{figure}[http!]
\centering
\includegraphics[scale=1]{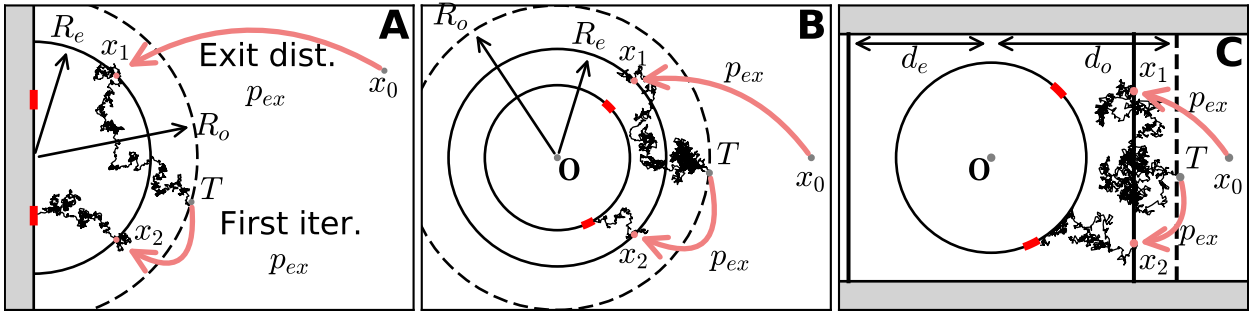}
\caption{(A) hybrid stochastic simulation procedure for two windows on the boundary of half-space. Brownian particles injected at $\x_0$ are directly place on semi-circle with radius $R_e$ according to the exit pdf $p_{HS}$ (red arrow). Inside the disk, Trajectories are generated by the Euler's scheme \ref{euler} until it passe outside the radius $R_o > R_e$, where the trajectory is terminated at point $T$ and restarted at a new position determined by the pdf $p_{HS}$ . (B) same as in (A) but for a ball. (C) hybrid simulation  scheme for windows on a disk in a strip. Brownian particles are injected at the boundary $x=d_e$ based on the exit probability distribution $p_S$. Trajectories with $x>d_o$ or $x<-d_o$ are re-injected at $x=\pm d_e$ according to $p_S$ (same procedure as in (A)).}
\label{fig:simulation}
\end{figure}

\begin{enumerate}
\item The first step consists of replacing Brownian paths by repositioning a Brownian particle to the boundary of an imaginary circle $C_i$ with radius $R_e$ (Fig. \ref{fig:simulation}A-B). The position of the particle on $C_i$ is computed from the exit distribution $p_{ex}$ of the steady-state Fokker-Planck equation with zero absorbing boundary condition on $C_i$. The exit point probability $p_{ex}$ is actually here the Green's function of the Laplace operator with zero absorbing boundary condition on $C_{i}$.
\item In the second step, we define a larger disk $D(R_0)$ of radius $R_0>R_e$ and run Brownian trajectories in the domain $D(R_e)-D(R)$ after starting on $C_{i}$, until they are either terminated because they escape the disk $D(R_e)$ (point T in fig \ref{fig:simulation}) or are absorbed on a small window. When a Brownian particle escapes through $\p D(R_e)$, we resume the trajectories by assigning a new random initial position on $C_{i}$ choosing uniformly distributed according to the exit probability $p_{ex}$.
\end{enumerate}
We present below the explicit Green's function and the steady-state flux for different geometries of Fig. \ref{fig:simulation}A-C. Because the splitting probability between the windows does not depend on the arrival time, we did not take into account the temporal aspect in the renewal process, when re-injecting Brownian particles.

For part (1), we need to use the explicit Green's function for the given geometry. We consider here three examples: half-disk on the boundary of the two-dimensional half-plane, the exterior of a disk in two dimensions and the two-dimensional half-strip. For part (2), we simulate particle trajectories using the Euler's scheme for the position $\x(t)$
\beq
  \label{euler}
  \x(t+\Delta t)=\x(t)+\sqrt{2D\Delta t}\w,
\eeq
where $\w$ is a two-dimensional normal distributed vector with zero
mean and variance one and $D$ the diffusion coefficient. The time step $\Delta t$ is chosen such that the mean square displacement between two time points is smaller than the size of the absorbing window $\eps$.
\subsection{ Construction of the mapping using explicit Green's functions} \label{sec:greens-funct-halfspace}
The first part of the stochastic-analytic hybrid algorithm consist in mapping the source to a point on the artificial circumference $C_i$. The construction starts with the explicit exit distribution $p_{ex}$ of the Laplace operator with zero absorbing boundary condition on $C_i$. The initial point $x_1$ is chosen randomly distributed according to the probability $p_{ex}(s)$, where $s$ is the arclength coordinate.
\subsubsection{Hybrid map positioning  for the full space} \label{sec:full}
We start with the explicit external Neumann-Green's function in $\rR^{2}$ with zero absorbing boundary condition on a disk $D(R)$ of radius R.
\beq
\begin{aligned}
\label{green3_full}
-\Delta_{\y} G(\x,\y) & = \delta(\x-\y), &\quad &\text{for}\;\; \x, \y\, \in \,\rR^{2}, \\
 G(\x,\y) & =0 &\quad &\text{for}\;\; \y\, \in\, \p B \cap \rR^{2},\ \x \in \,\rR^{2}.
\end{aligned}
\eeq
The solution is constructed by the method of images \cite{Melnikov} and given by
\beq
\label{greenfct-disk}
G(\x,\y)=-\frac{1}{2\pi}\left(\ln|\x-\y|-\ln\left|\x-\frac{R^2}{|\y|^2}\y\right|-\ln\frac{|\y|}{R}\right).
\eeq
Thus the probability distribution of exit points $p_{ex}$ on the boundary $\p D(R)$ given that source is located at position $\x_0$ is computed  by normalizing the flux \cite{DSP},
\beq
p_a(\y|\x_0)=\ds \frac{\ds \frac{\p G }{\p \n_{y}}(\y,\x_0)}{\ds \oint_{\p D(R)}\frac{\p G }{\p \n_{y}}(\y,\x_0)dS_{\y}},
\eeq
The flux is computed in polar coordinates $r=|\x|$, $\rho=|\y|$ and the angles $\theta$ and $\theta'$ (with the horizontal axis) of points $\x$ and $\y$ respectively:
\beq
\label{flux-disk}
p_{ex}(r,\theta;\theta')=R\frac{\p G}{\p
  \rho}\Bigl|_{\rho=R}=\ds \frac{1}{2\pi}\frac{\ds \frac{r^2}{R^2}-1}{\ds \frac{r^2}{R^2}-2\frac{r}{R}\cos(\theta-\theta')+1}.
\eeq
Note that indeed $\int_{\p D(R)}\frac{\p G }{\p \n_{y}}(\x,\y)dS_{\y}=1$.
The probability \ref{flux-disk} is used to computed the position of the sequence of points $\x_1, \x_2,..$ randomly and uniformly chosen, until the trajectory is finally absorbed at one of the windows. Each time a trajectory hits the external circle of radius $R_0$, the motion is immediately resumed at one of the points $x_i$ ($i=1..$). This procedure disregards the absolute time of the trajectories.

\subsubsection{Hybrid map positioning  for a half-space $\rR^{2}_+$} \label{sec:halfspace}
The Neumann-Green's function $G_{HS}$ for the half-space $\rR^{2}_+$ with zero absorbing boundary condition on a half a disk of radius R is the solution of the boundary value problem
\beq \begin{aligned}
\label{green3_halfspace}
-\Delta_{\y} G_{HS}(\x,\y) & = \delta(\x-\y), &\quad &\text{for}\;\; \x, \y\, \in \,\rR^{2}_+, \\
 \frac{\p G_{HS}}{\p n_{\y}}(\x,\y) & =  0, &\quad &\text{for}\;\; \y\, \in\, \p\rR^{2}_+,\ \x \in \,\rR^{2}_+,\\
 G_{HS}(\x,\y) & =0 &\quad &\text{for}\;\; \y\, \in\, \p B \cap \rR^{2}_+,\ \x \in \,\rR^{2}_+,
\end{aligned}
\eeq
The solution is obtained by the method of image charges using the Green's function for the absorbing disk in free space $G(\x,\y)$ computed in section in eq. \ref{sec:full}. The Green's function for the half-space $\rR^{2}_+$ is then constructed by symmetrizing with respect to the reflecting z-axis:
\beq
\label{greenfct-halfspace}
G_{HS}(\x,\y) & =& \frac{1}{2}[G(\x,\y)+G(x,\tilde \y)] -\frac{1}{4\pi} \Bigl( \left(\ln|\x-\y|-\ln\left|\x-\frac{R^2}{|\y|^2}\y\right|-\ln\frac{|\y|}{R}\right)+\\
&&\left(\ln|\x-\tilde\y|-\ln\left|\x-\frac{R^2}{|\tilde\y|^2}\tilde\y\right|-\ln\frac{|\tilde\y|}{R}\right)\Bigr), \nonumber
\eeq
where $\tilde \y$ is the mirror reflection of $\y$ on the vertical axis. The exit probability distribution is the flux through the absorbing half disk boundary
\beq \label{flux-halfspace}
\begin{aligned}
p_{ex}(r,\theta;\theta')=2R\ds\frac{\partial G}{\partial \rho}\Bigl|_{\rho=R}
                         =\ds\frac{\ds\frac{r^2}{R^2}-1}{2\pi}&\left[\frac{1}{\ds 1-2\frac{r}{R}\cos(\theta-\theta')+\frac{r^2}{R^2}}\right.\\ &\;\;+\left.\frac{1}{\ds 1+2\frac{r}{R}\cos(\theta+\theta')+\frac{r^2}{R^2}}\right],
\end{aligned}
\eeq
where the length in polar coordinates are $r=|\x|$, $\rho=|\y|$ and
 the angles $\theta$ and $\theta'$ of $\x$ and $\y$ are given with respect to the horizontal axis respectively.

\subsubsection{Green's function for the semi-strip} \label{sec:greens-function-semistrip}
Finally, we summarize here the Neumann-Green's function $G_{Se}$ for a semi-strip
\beq
\Omega_a=\{(x_1,x_2)\in \rR^2 |x_1>0,0<x_2<a\},
\eeq
of width $a>0$. The normalized flux is the distribution of exit points \cite{DSP}. A zero absorbing boundary condition is imposed on the boundary $\p\Omega_1=\{(0,x_2)|0<x_2<a\}$ and a reflecting boundary condition on the rest of the strip $\p\Omega_2=\{(x_1,0)|x_1>0\}\cup\{(x_1,a)|x_1>0\}$ (Fig. \ref{fig:simulation}C). The function $G_{Se}$ is solution of the boundary value problem
\beq
\begin{aligned}
\label{green-semistrip}
-\Delta_{\y} G_{Se}(\x,\y) & = \delta(\x-\y), &\quad &\text{for}\;\; \x, \y\, \in \,\Omega, \\
 \frac{\p G_{Se}}{\p n_{\y}}(\x,\y) & =  0, &\quad &\text{for}\;\; \y\, \in\, \p\Omega_2,\ \x \in \,\Omega,\\
 G_{Se}(\x,\y) & =0 &\quad &\text{for}\;\; \y\, \in\, \p\Omega_1,\ \x \in \,\Omega.
\end{aligned}
\eeq
The exit probability distribution $p_{ex}(x_2;y_1,y_2)$ is given explicitly (see appendix) by the flux through the artificial boundary $\p\Omega_1$
\begin{equation}
  \label{eq:strip-y-distribution}
  p_{ex}(x_2;y_1,y_2)=\frac{\p G_{Se}}{\p y_1}\Bigl|_{y_1=0}=\frac{\sinh\omega y_1}{2a}\Bigl[\frac{1}{\cosh\omega y_1-\cos\omega(x_2+y_2)}+\frac{1}{\cosh\omega y_1-\cos\omega(x_2-y_2)}\Bigr]\,.
\end{equation}
\section{Results of the hybrid algorithm}
\subsection{Computing the splitting probability in half a space}
\begin{figure}[http!]
\centering
\includegraphics[scale=1]{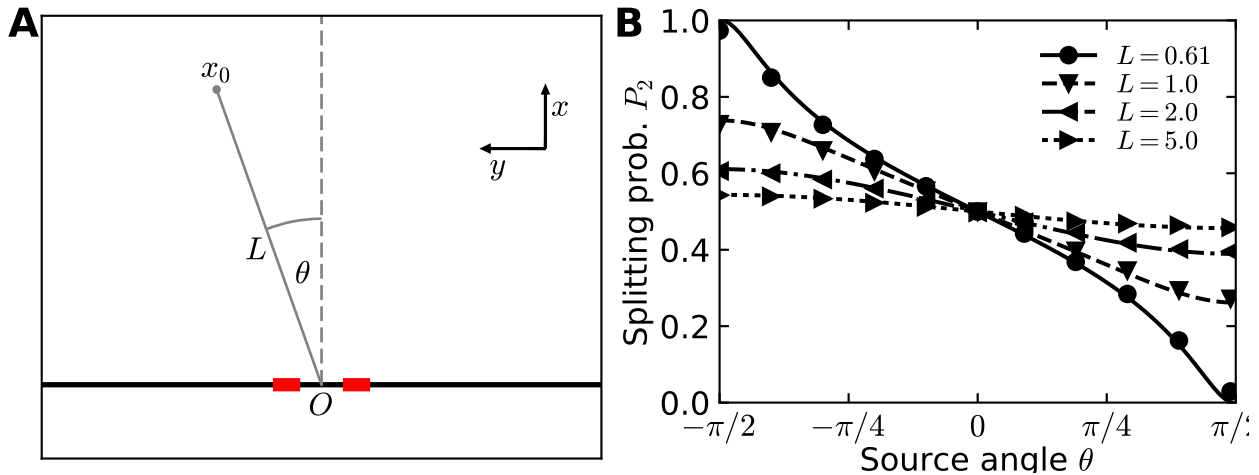}
\caption{Diffusion fluxes to small windows on the boundary of a half-plane. (A) Particles are released at the source $\x_0$ at a distance $L=|\x_0|$ from the origin forming an angle $\theta$ with the $x$-axis. (B) Splitting probability (normalized flux) at window 2 as a function of $\theta$ for different $L$. {The exact analytical solution given in equation~\eqref{eq:p2-halfplane2} (lines) is compared to hybrid simulations (markers).}}
\label{Fig-result}
\end{figure}

To illustrate the mixed stochastic-analytical algorithm described in the previous section, we computed the splitting probability for two windows located on the y-axis, where Brownian particles are released at position $\x_0$ in the half space (Fig. \ref{Fig-result}A). We compare the splitting probability computed analytically (formula \ref{eq:p2-halfplane}) with the results of the hybrid stochastic-analytical simulations and found perfect agreement (Fig. \ref{Fig-result}B) when we vary the distance of the source $L=|\x_0|$ from the origin $O$.
When the source $\x_0$ is located more than 10 times the distance $d=|\x_1-\x_2|$ between the two windows, the difference in fluxes between the two windows decays to less than $5\%$, suggesting that small fluctuations of the same order of magnitude render it impossible to measure the direction of the source from the steady state fluxes. We obtained similar results for the case of the disk, but not for a disk located in a band with reflecting walls, where the detection sensitivity extends much further \cite{UliarXiv}.

{In order to study the sensitivity with respect to relative position of the windows and the source, we previously introduced the sensitivity ratio as the difference between the fluxes to the two windows \cite{UliarXiv}
\beq\label{Ratio}
r(\x_1,\x_2,\x_0)=|P_1(\x_1,\x_2,\x_0)-P_2(\x_1,\x_2,\x_0)|\,.
\eeq
We then used this ratio to define the domain of sensitivity as consisting of all possible source locations that yield a ratio $r(\x_1,\x_2,\x_0)$ larger than a pre-defined threshold $T$. Similarly, we defined the maximum detection threshold function as \cite{UliarXiv}
\beq
f(\x_0)=\max_{\x_1,\x_2}r(\x_1,\x_2,\x_0)\,.
\eeq
For a disk of radius $R$ in free space, this maximum is indeed obtained for a window configuration aligned with the position of the source and symmetric with respect to the center of the disk centered at the origin. Hence, we obtain $\x_2=-\x_1$ and $|\x_1|=|\x_2|=R$ and
 \beq
f(\x_0)=&\frac{1}{2}\frac{\ln \frac{|\x_1-\x_0|\left|\x_1-\bar{\x_0}\right|}{|\x_2-\x_0|\left|\x_2-\bar{\x_0}\right|}}{\{\log|\x_1-\x_2|-\log \eps\}}\\
       =&\ds \frac{1}{2}\frac{\ds \ln \frac{\left|R-|\x_0|\right| \left|R-R^2/|\x_0|\right|}{\ds\left|R+|\x_0|\right| \left|R+R^2/|\x_0|\right|}}{\{\log|2R|-\log \eps\}}\\
       =&\ds \frac{1}{2}\frac{\ds \ln \frac{\left|1-|\x_0|/R\right| \left|1-R/|\x_0|\right|}{\ds\left|1+|\x_0|/R\right| \left|1+R/|\x_0|\right|}}{\{\log|2R|-\log \eps\}}.
\eeq
In particular, a Taylor expansion of $f(\x_0)$ for large source position $L=|x_0|$, leads to the decay of the maximum detection threshold function \cite{UliarXiv}
\beq \label{maxdetectionthreshold}
f(\x_0)= \ds\frac{2R}{L\log\frac{2R}{\eps}}+o\left(\frac{1}{L}\right).
\eeq

For two windows located on the boundary of half-space, the sensitivity ratio \ref{Ratio} can only be influenced by the spacing between the windows $d=|\x_1-\x_2|$. Therefore, we do not need to find the optimal arrangement and can directly compute the sensitivity ratio
\beq
r(d, L, \theta)= \left|\frac{1}{2}\frac{1}{\ln(d/\eps)}\ln\left[\frac{\frac{d^2}{4}+L^2-Ld\sin{\theta}}{\frac{d^2}{4}+L^2+Ld\sin{\theta}}\right]\right|\,,
\eeq
where $\theta$ is the angle between the $x$-axis and the vector from the origin $O$ to the source location $x_0$. A Taylor expansion for $L>>d$ of the logarithmic term yields to
\beq
r(d, L, \theta)=\frac{d}{L}\frac{|\sin\theta|}{\ln(d/\eps)}+o\left(\frac{d}{L}\right),
\eeq
where the maximum of the detection threshold is similar to the one of the disk in equation \ref{maxdetectionthreshold} with $d=2R$ and $\theta=\pm \pi/2$.

We conclude that for a disk and the half-plane, the detection threshold decays algebraically with the distance. In a biological context, which involve two different types of absorbing windows, each accepting only one of two types of Brownian particles, the splitting probabilities are independent. In this case, we would define the sensitivity as the product of each particle's sensitivity function by
\beq
f_{\text{2 classes}}(\x_0)= f(\x_0)^2 \propto \ds\left(\frac{d}{L\ln(d/\eps)}\right)^2+o\left(\frac{1}{L^4}\right).
\eeq
}
Interestingly, this formula would predict for that case a decay of the splitting probability of $1/dist^2$ with respect to the source position.
\subsection{Recovering the position of the source from the fluxes to several windows} \label{sec:3receptod}
To reconstruct the location of a source from the measured fluxes, at least three windows are needed. Indeed, with two windows only, a source located on the line perpendicular to the one of the connecting windows would, for example generate the same splitting probability $P_1=P_2$, leading to a one dimensional curve degeneracy for the reconstructed source positions $\x_0$.

\begin{figure}[http!]
\centering
\includegraphics[scale=1]{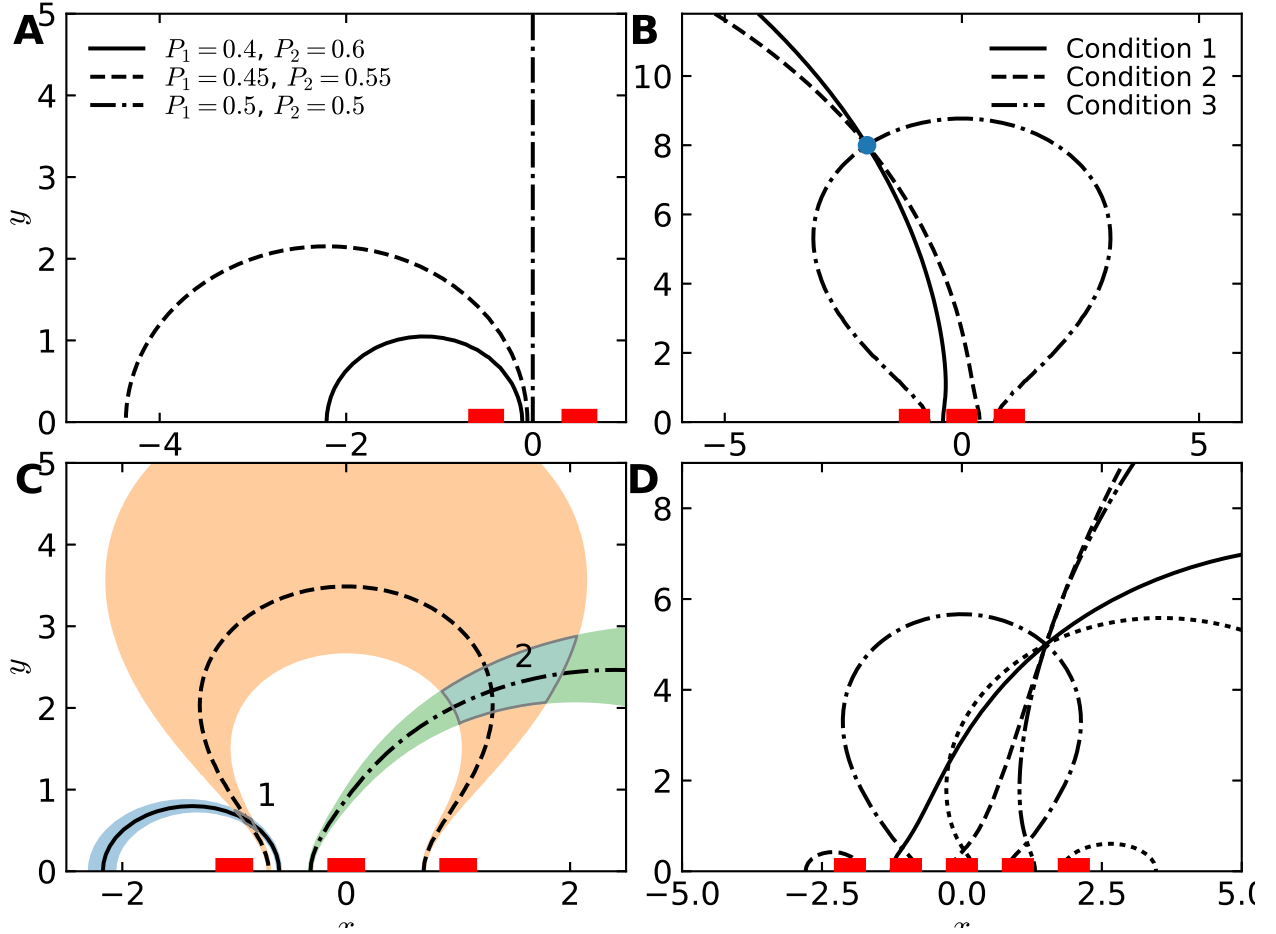}
\caption{Reconstruction of the source position in the half-plane. (A) Two windows placed a distance $d=1$ apart allow the recovery of the source position up to a curve. Three different flux configurations are shown. (B) Three windows positioned at $y=-1, 0, 1$ yield two independent curves, the intersection of which is the position of the source {at $\x_0=(-2, 8)$}. The redundant third condition is shown for completeness. (C) The shaded areas indicate the uncertainty resulting from fluctuations in the fluxes with an amplitude of $\eta=0.005$. The resulting sensitivity of the reconstructed source position (overlapping shaded areas 1 and 2) is highly inhomogeneous. (D) Five windows with nearest-neighbor distance $d=1$ yield four independent curves. The point of intersection of all curves is the recovered source position.}
\label{Fig3}
\end{figure}

To study the reconstruction of a source location $\x_0$ from the splitting probabilities, we need to invert system \ref{cond_many}.  The general solution is given by
\beq\label{gexp2}
P(\x,\x_0)=&G(\x,\x_0)  +A_1 \{\log|\x-\x_1|-\log \eps\}+A_2 \{\log|\x-\x_2|-
\log \eps\}\\
\nonumber
&+A_3 \{\log|\x-\x_3|-\log \eps\} +C,
\eeq
where $A_1,A_2,A_3,C$ are constants to be determined. Following the step of section \ref{manywindows}, the three absorbing boundary conditions for $P(\x,\x_0)$ give
\beq\label{cond2}
G(\x_1,\x_0)  +A_2 \{\log|\x_1-\x_2|-\log \eps\}+A_3 \{\log|\x_1-\x_3|-\log \eps\} +C=0\\
G(\x_2,\x_0)  +A_1 \{\log|\x_2-\x_1|-\log \eps\} +A_3 \{\log|\x_2-\x_3|-\log \eps\}+C=0\\
G(\x_3,\x_0)  +A_1 \{\log|\x_1-\x_3|-\log \eps\}+A_2 \{\log|\x_2-\x_3|-\log \eps\} +C=0.
\eeq
The normalization condition for the fluxes is
\beq \label{compatibility3}
\pi A_1+\pi A_2+\pi A_3=1
\eeq
and the solution is
\beq\label{cond4}
G(\x_1,\x_0)-G(\x_2,\x_0)-\frac{1}{\pi} \log\frac{|\x_2-\x_1|}{\eps} +2 A_2 \log\frac{|\x_2-\x_1|}{\eps}+A_3 \log\frac{|\x_1-\x_3||\x_2-\x_1|}{|\x_2-\x_3|\eps} =0\\
G(\x_1,\x_0)-G(\x_3,\x_0)-\frac{1}{\pi}\log\frac{|\x_3-\x_1|}{\eps} +A_2 \log\frac{|\x_1-\x_2||\x_3-\x_1|}{|\x_2-\x_3|\eps}+2A_3 \log\frac{|\x_3-\x_1|}{\eps} =0.
\eeq
Using the determinant:
\beq\label{conddelta}
\Delta_{123}=\left(\log\frac{d_{13}d_{12}}{d_{32}\eps}\right)^2-4\log \frac{d_{12}}{\eps} \log \frac{d_{13}}{\eps},
\eeq
and the general notation for any i,j
\beq
d_{ij}=|\x_i-\x_j|,
\eeq
we get
\beq\label{a22}
A_2=\frac{  \log\frac{d_{13}d_{12}}{d_{32}\eps}(G_{30}-G_{10} +\frac{1}{\pi} \log\frac{d_{13}}{\eps})-( G_{1 0}-G_{2 0}+\frac{1}{\pi}\log\frac{d_{12}}{\eps})  \log\frac{d_{13}^2}{\eps^2  })}{\Delta_{123}}.
%
\eeq
\beq\label{a33}
A_3=\frac{  \log\frac{d_{13}d_{12}}{d_{32}\eps}(G_{20}-G_{10} +\frac{1}{\pi} \log\frac{d_{12}}{\eps})-( G_{1 0}-G_{3 0}+\frac{1}{\pi}\log\frac{d_{13}}{\eps})  \log\frac{d_{12}^2}{\eps^2  })}{\Delta_{123}}
%
\eeq
and
\beq\label{a11}
A_1=\frac{1}{\pi}-A_2 -A_3.
\eeq
This equation resolve uniquely the problem of determining the source location from the fluxes. Indeed, choosing $\alpha>0$ and $\beta>0$ such that $\alpha+\beta<1$, the position of the source is located at the intersection of the two curves:
\beq \label{compatibility1}
\alpha&=&\pi A_1=\int_{\p\Omega_{1}} \frac{\p P(\x,\y))}{\p \n}dS_{\x}\\
\nonumber
\beta&=&\pi A_2=\int_{\p\Omega_{2}} \frac{\p P(\x,\y))}{\p \n}dS_{\x}.
\eeq
Due to the normalization condition \ref{compatibility}, the flux condition (relation \ref{a11}) on window 3 is redundant. Solving analytically the system \ref{compatibility1} remains difficult, hence we investigate the position of the source $\x_0$ numerically by inverting system \ref{compatibility1} using expression \ref{a22}-\ref{a33}. The result is shown in figure \ref{Fig3}B. We conclude that for three and more receptors, it is always possible to reconstruct the source location.
We positioned here the source 8 times the distance between the effective receptors, that could represent in reality clusters of receptors. Indeed, this organization would correspond to receptors located on the diameter of a round cell. So we interpret the present result as recovering a source located 8 times the diameter of a cell (see for an application \cite{UliarXiv}).

We next tested the effect of possible uncertainty in the steady state fluxes on the recovery of the source $\x_0$, by adding a small perturbation to the fluxes, so that $\alpha=\alpha_0(1+\eta),\beta=\beta_0(1+\eta)$ with $\eta \ll 1$ in \ref{source}. Using numerical solutions, we found that the resulting uncertainty in $\x_0$ has a highly non-linear spatial dependency, as shown by the relative sizes of the areas labelled 1 and 2 in Fig. \ref{Fig3}C.

Finally, We studied the consequence of adding more windows. These additional windows allows to refine the reconstruction of the source. We increased the number to 5 (Fig. \ref{Fig3}D) and indeed found that the source is precisely located at the intersection of all curves for a given set of fluxes. There are other points at which two curves intersect, however, there is only one location where more than two curves (all of them) intersect, which corresponds to the source position. We conclude that having several windows could reduce the area of the uncertainty region when the fluxes contains some steady-state fluctuations.

\begin{figure}[http!]
\centering
\includegraphics[scale=1]{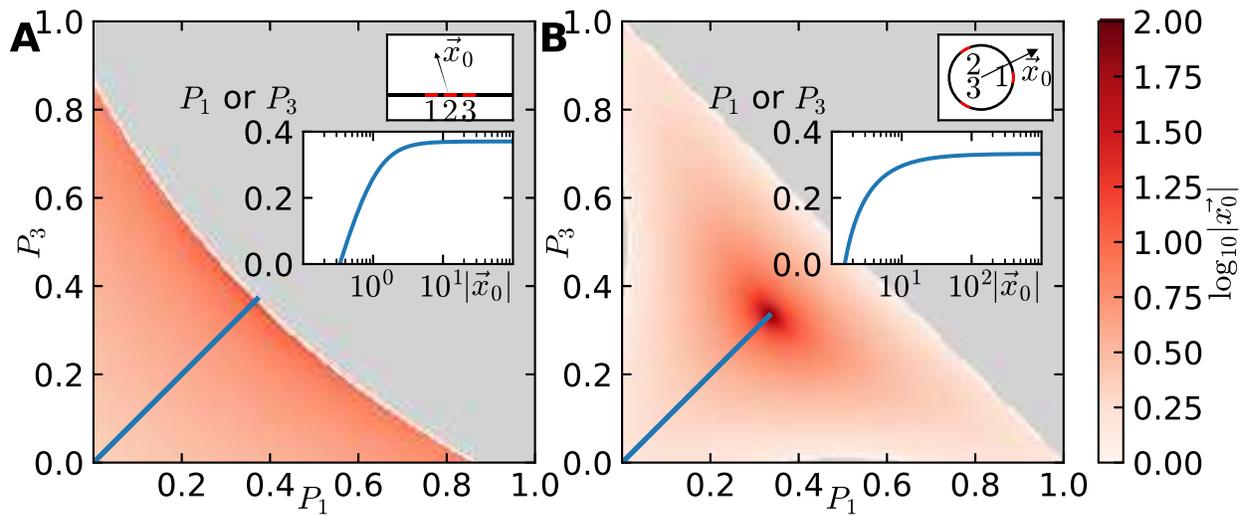}
\caption{Distance to the source as a function of the flux. (A) Three receptors arranged on the boundary of the half-plane, and (B) three receptors equally spaced on the circumference of a disk. The color shading indicates the logarithmically scaled distance to the source as a function of the two independent fluxes $P_1$ and $P_3$. The grey area indicates invalid flux combinations due to the condition $\sum_i P_i=1$. The top inset displays the arrangement of windows and the lower inset shows the relationship between the fluxes and the source distance along the blue line. Note that in both cases, the distance increases nonlinearly as a function of $P_{1/3}$, hence the recovery of the source position is robust against flux noise only at short distances.}
\label{Fig4}
\end{figure}

In order to further investigate the sensitivity of the recovered source position to small fluctuations in the flux, we numerically solved system \ref{sys1} for three windows. Figure \ref{Fig4}A shows how the recovered distance depends the fluxes $P_1$ and $P_3$ for windows on the boundary of the half-plane, while Fig. \ref{Fig4}B displays the same result for a disk in $\rR^2$.

\section{Discussion and conclusion}
In this paper, we studied the steady-state distribution of fluxes across absorbing windows located on a surface of a disk embedded in the plane and narrow strip and on the boundary of the half-plane. Interestingly, we found that with three absorbing windows, it is possible to recover the location of a Brownian source of particles from the splitting probability in two dimensions. The analytical computations are based on matched asymptotics to construct the Green's function used in the analysis. We developed here a new mixed efficient algorithm to compute numerically the fluxes by generating truncated Brownian trajectories. { Both the analytical results and the simulation procedure presented here rely on the Green's function of the domain of interest. The motivation of the present work is the problem that a cell has to face for navigation: finding a gradient source inside a tissue, which is often a two dimensional rather than three dimensional problem, as cells are moving along other cells, thus reducing the dimension. In addition, many chemotaxis experiments occurring in microfluidic chambers are almost two dimensional. The method we developed here are however applicable with no restriction in three dimensions. In particular, the same procedure can be used to simulate sensing in three dimensional spaces.

Other methods use Green's function for the simulations of stochastic particle trajectories (reaction-diffusion method~\cite{Zon2005}), applied to transient receptor binding~\cite{Kaizu2014}. However, these kinetic simulations is quite different from the direct time step propagation method that we have developed here. The model is a molecular gradient, generated by a fixed source emitting Brownian particles. We simplified here the cell geometry to a round disk containing small fast absorbing targets on its surface (receptors). A diffusing molecule may find one of these receptors, leading to its activation. We neglected the binding time. Furthermore, we only focus on the steady-state regime in which the external gradient is already established and thus there is no intrinsic time scale. This is in contrast to transient regime, that could represent the regime shortly after the source first starts to emit, in which the time scale can be defined as the first passage time of particles to reach the cell.

Receptor activation can mediate cellular transduction that transform an external environment signal into a cellular biochemical activation cascade. When a cell has to differentially compare the flux from one side and the other, the local transduction of the signal at the scale of a receptor must not be homogenized throughout the rest of the cell domain, such that the local information about the gradient directionality is preserved. The internal transduced signal can be carried by the concentration of second messenger or diffusing surface molecules. Hence, receptor activations need to be localized inside the cell, leading to an asymmetrical response. Therefore, we studied here the flux to stationary and localized receptors, and do not replace receptors with a homogenized boundary condition that is unable to preserve flux differences across cells and directional information.

Estimating the fluctuations in the number of receptor-ligand molecules reaching a cell can be found in \cite{BergPurcell,Goodhill1,Berezhkovskii2013}. These models are generally based on homogenization of the boundary condition, rendering it impossible to recover any directional information since they assign the same flux to the entire boundary. In order to find the fluxes to each window separately, we based our analysis on the narrow escape theory \cite{holcmanschuss2015}. Interestingly, we find that in two dimensions, the difference in the probability flux decays algebraically with $1/L$, where $L$ is the distance to the source. Although we focus the present investigation only on two or three windows here, the results would be very similar for clusters of windows \cite{JPA-holes}. The low number of relevant windows located on a neurite must be involved in detecting a gradient concentration, its direction to turn, its forward or retracting motion \cite{BouziguesPLoS1,Stettler}. The model we have studied here is equivalent to fast binding \cite{Kaupp1} (without rebinding). It remains a challenge to explain how bacteria \cite{Heinrich3}, sperm \cite{Kaupp2} or neurite growth \cite{Steller} localize the source of a chemotactic gradient. Sensing the fluxes to receptor across the cell body is certainly the first step and the present study shows that this information is sufficient to reconstruct the location of a source.  The optimal distribution of receptors could also vary from a uniform to a cluster distribution of receptors, a problem that should also be studied.
\section{Appendix}\label{appendix1}
\subsection{Stability of the hybrid-simulations}
We evaluated the stability of the computational method described in section \ref{s:algo} by varying the inner radius $R_{e}$ where Brownian particles are injected. The scheme of the algorithm is presented in Fig. \ref{FigA1}A. Varying the inner radius from $R_{e}=1.3$ to $3$ has no impact on the measured steady-state fluxes, as shown in Fig. \ref{FigA1}B. This method neglects the return of far away trajectories, which in principle occurs with probability 1 in dimension 2, due to the recurrent properties of the Brownian motion.
\begin{figure}[http!]
\centering
\includegraphics[scale=1]{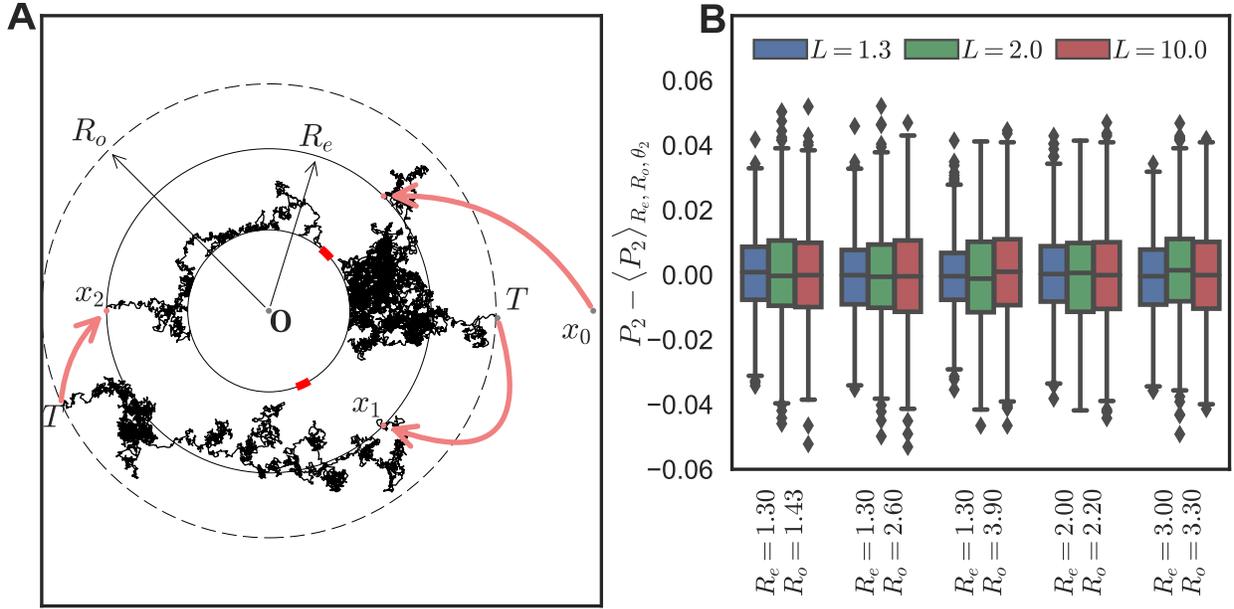}
\caption{ Simulation results are independent of the choice of entering radius $R_e$ and outer radius $R_o$. (A) Two windows are placed on a disk. Brownian particles originating from the source at $x_0$ are injected at the circumference of the circle with radius $R_e$. Trajectories leaving the region with radius $R_o$ are restarted at a radius $R_e$. (B) Varying the exit radius $R_e$ and $R_o$ does not change the splitting probability (flux) from simulations. The box plot shows the distribution of errors in the flux to window 2. The error is defined as the deviation of the flux from the mean over all observed radii combinations $(R_e, R_o)$ and all angles $\theta_2$, but separately for all source distances $L$. Note that there are no systematic deviations when changing either $R_e$ or $R_o$.}
\label{FigA1}
\end{figure}
\subsection{Explicit Green's function in a band} \label{sec:band}
The hybrid algorithm is based on the exact expression of the Neumann-Green's function $G_{Se}$ for the semi-strip
\beq
\Omega_a=\{(x_1,x_2)\in \rR^2 |x_1>0,0<x_2<a\}
\eeq
where $a>0$. The normalized flux is the distribution of exit points \cite{DSP}.
We impose zero absorbing boundary condition on the boundary $\p\Omega_1=\{(0,x_2)|0<x_2<a\}$ and reflecting boundary condition on the rest of the strip $\p\Omega_2=\{(x_1,0)|x_1>0\}\cup\{(x_1,a)|x_1>0\}$. The boundary value problem is
\beq
\begin{aligned}
\label{green-semistrip_appdx}
-\Delta_{\y} G_{Se}(\x,\y) & = \delta(\x-\y), &\quad &\text{for}\;\; \x, \y\, \in \,\Omega, \\
 \frac{\p G_{Se}}{\p n_{\y}}(\x,\y) & =  0, &\quad &\text{for}\;\; \y\, \in\, \p\Omega_2,\ \x \in \,\Omega,\\
 G_{Se}(\x,\y) & =0 &\quad &\text{for}\;\; \y\, \in\, \p\Omega_1,\ \x \in \,\Omega.
\end{aligned}
\eeq
We compute $G_{Se}(\x,\y)$ by expanding in eigenfunction following the classical method of \cite{Melnikov} (p. 80). To start we write the following Ansatz
\[
u(x_1,x_2)=\int_0^\infty\int_{0}^{a}G_{Se}(x_1,x_2;y_1,y_2)f(y_1,y_2)dy_1 dy_2,
\]
with $x_1$, $x_2$ the components of $\x$ and $y_1$, $y_2$ the components of
$\y$, which solves the inhomogeneous diffusion equation
\beq
\begin{aligned}
\label{udiffeq}
-\left(\frac{\p^2}{\p x_1^2}+\frac{\p^2}{\p x_2^2}\right) u(x_1,x_2) & = f(x_1,x_2), &\quad &\text{for}\;\; x_1>0\,,\; 0 < x_2 < a, \\
 \frac{\p u}{\p x_2}(x_1,x_2) & =  0, &\quad &\text{for}\;\; x_1>0\,,\;\; x_2=0 \; \text{and}\; x_2=a,\\
 u(0,x_2) & =0, &\quad &\text{for}\;\; 0<x_2<a.
\end{aligned}
\eeq
Because $x_2$ is bounded to between $0$ and $a$, we can write $u$ and $f$ in
terms of a Fourier series along $x_2$ \beq
\label{ufourier}
u(x_1,x_2)=\sum_{n=0}^{\infty}u_n(x_1) \cos\omega_nx_2\,,\quad\omega_n=\frac{n\pi}{a},
\eeq
\beq
\label{ffourier}
f(x_1,x_2)=\sum_{n=0}^{\infty}f_n(x_1) \cos\omega_nx_2\,,\qquad f_n(x_1)=\frac{2}{a}\int_0^af(x_1,x_2)\cos(\omega_n x_2)dx_2,
\eeq
with $\omega_n=\frac{n\pi}{a}$ By inserting this expression for $u$ into equation~(\ref{udiffeq}) we arrive at the following ODE for the $u_n$
\[
u_n''-\omega_n^2u_n=f_n\,.
\]
For $n=0$, the fundamental solutions to the homogeneous equation $u_0''=0$ are $u_0^{(1)}=1$ and $u_0^{(2)}=x_1$. Therefore, the inhomogeneous problem is solved by
\[
u_0(x_1)=\int_0^{x_1}y_1f_0(y_1)dy_1+C_1+x_1\Bigl(-\int_0^{x_1}f_0(y_1)dy_1+C_2\Bigr)=\int_0^\infty\min\{x_1,y_1\}f_0(y_1)dy_1,
\]
where $C_2=\int_0^{\infty}f(y_1)dy_1$ due to the boundedness condition on the solution as $x_1\to 0$ and $C_1=0$ due to the absorbing boundary at $x_1=0$.

For $n>1$, the fundamental solutions to the homogeneous system $u_n''=\omega_n^2 u_n$ are given by $u_n=\exp(\pm\omega_n x_1)$. Hence,
\[
\begin{aligned}
u_n(x_1)&=\frac{e^{\omega_nx_1}}{2\omega_n}\Bigl(-\int_0^{x_1}e^{-\omega_ny_1}f_n(y_1)dy_1+C_1\Bigr)-\frac{e^{-\omega_nx_1}}{2\omega_n}\Bigl(\int_0^{x_1}e^{\omega_ny_1}f_n(y_1)dy_1+C_2\Bigr) \\
&=\frac{e^{\omega_nx_1}}{2\omega_n}\int_{x_1}^{\infty}e^{-\omega_ny_1}f_n(y_1)dy_1+\frac{e^{-\omega_nx_1}}{2\omega_n}\Bigl(\int_0^{x_1}e^{\omega_ny_1}f_n(y_1)dy_1-\int_0^\infty e^{-\omega_ny_1}f_n(y_1)dy_1\Bigr) \\
&=\frac{1}{2\omega_n}\int_{0}^{\infty}\Bigl(e^{-\omega_n|x_1-y_1|}-e^{-\omega_n(x_1+y_1)}\Bigr)f_n(y_1)dy_1
\end{aligned}
\]
Thus, the complete solution to equation~(\ref{udiffeq}) reads
\[
\begin{aligned}
u&(x_1,x_2)=\int_0^\infty\Bigl(\min\{x_1,y_1\}f_0(y_1)+\frac{a}{2\pi}\sum_{n=1}^{\infty}\frac{1}{n}\bigl[e^{-\omega_n(y_1+x_1)}-e^{-\omega_n|y_1-x_1|}\bigr]f_n(y_1)\cos\omega_nx_2\Bigr)dy_1 \\
&=\int_0^a\int_0^\infty\Bigl(\frac{1}{a}\min\{x_1,y_1\}+\frac{1}{\pi}\sum_{n=1}^{\infty}\frac{1}{n}\bigl[e^{-\omega_n(y_1+x_1)}-e^{-\omega_n|y_1-x_1|}\bigr]\cos\omega_nx_2\cos\omega_ny_2\Bigr)f(y_1,y_2)dy_1dy_2\\
&=\int_0^a\int_0^\infty\Bigl(\frac{1}{a}\min\{x_1,y_1\}+\\
&\qquad+\frac{1}{2\pi}\sum_{n=1}^{\infty}\frac{1}{n}\bigl[e^{-\omega_n(y_1+x_1)}-e^{-\omega_n|y_1-x_1|}\bigr]\bigl[\cos\omega_n(x_2-y_2)+\cos\omega_n(x_2+y_2)\bigr]\Bigr)f(y_1,y_2)dy_1dy_2,
\end{aligned}
\]
where we inserted the fourier coefficients for $f$ from
equation~(\ref{ffourier}). By inspection, we arrive at the expression for the Green's function
\[
G_{Se}(x_1,x_2;y_1,y_2)=\frac{1}{a}\min\{x_1,y_1\}+\frac{1}{2\pi}\sum_{n=1}^{\infty}\frac{1}{n}\bigl[e^{-\omega_n(y_1+x_1)}-e^{-\omega_n|y_1-x_1|}\bigr]\bigl[\cos\omega_n(x_2-y_2)+\cos\omega_n(x_2+y_2)\bigr].
\]
Using the identity~(see \cite{Melnikov} p. 84)
\[
\sum_{n=1}^{\infty}\frac{q^{n}}{n}\cos n\phi=-\frac{1}{2}\ln(1-2q\cos\phi+q^2),
\]
we can further simplify to get
\beq
\label{greensfct-semistrip}
\begin{aligned}
  G_{Se}(x_1,x_2;y_1,y_2)=-\frac{1}{4\pi}\bigl[&\ln(1-2e^{-\omega|x_1-y_1|}\cos\omega(x_2+y_2)+e^{-2\omega|x_1-y_1|}) \\
  &+\ln(1-2e^{-\omega|x_1-y_1|}\cos\omega(x_2-y_2)+e^{-2\omega|x_1-y_1|}) \\
  &-\ln(1-2e^{-\omega(x_1+y_1)}\cos\omega(x_2+y_2)+e^{-2\omega(x_1+y_1)}) \\
  &-\ln(1-2e^{-\omega(x_1+y_1)}\cos\omega(x_2-y_2)+e^{-2\omega(x_1+y_1)}) \\
  &+\frac{4\pi}{a}\min\{x_1,y_1\}\bigr]\,.
\end{aligned}
\eeq
with $\omega=\pi/(2a)$. The exit
probability distribution is again given by the flux through the $\p\Omega_1$
boundary
\begin{equation}
  \label{eq:strip-y-distribution_appdix}
  p_{ex}(x_2;y_1,y_2)=\frac{\p G_{Se}}{\p y_1}\Bigl|_{y_1=0}=\frac{\sinh\omega y_1}{2a}\Bigl[\frac{1}{\cosh\omega y_1-\cos\omega(x_2+y_2)}+\frac{1}{\cosh\omega y_1-\cos\omega(x_2-y_2)}\Bigr]\,.
\end{equation}

\end{document}